\documentclass{aastex63}
\usepackage[utf8]{inputenc}

\begin{document}

\title{Investigating Possible Binarity for GJ 229B}

\author[0000-0002-4884-7150]{Alex R. Howe}
\affiliation{NASA Goddard Space Flight Center, 8800 Greenbelt Rd, Greenbelt, MD 20771, USA}
\affiliation{Center for Research and Exploration in Space Science and Technology, NASA/GSFC, Greenbelt, MD 20771}
\affiliation{Southeastern Universities Research Association, 1201 New York Avenue NW, Suite 430, Washington, DC 20005}

\author[0000-0002-8119-3355]{Avi M. Mandell}
\affiliation{NASA Goddard Space Flight Center, 8800 Greenbelt Rd, Greenbelt, MD 20771, USA}
\affiliation{GSFC Sellers Exoplanet Environments Collaboration}

\author[0000-0003-0241-8956]{Michael W. McElwain}
\affiliation{NASA Goddard Space Flight Center, 8800 Greenbelt Rd, Greenbelt, MD 20771, USA}

\submitjournal{ApJ Letters}

\begin{abstract}

GJ 229B, the first type-T brown dwarf to be discovered, has presented a tension between comparisons with evolutionary models and the larger-than-expected mass and radius values derived from spectroscopic and astrometric observations. We examine the hypothesis that GJ 229B is actually a binary sub-stellar object by using two grid-based fits using evolutionary models to explore the range of mass ratios of the possible binary components. We find that the best-fit component values are most consistent with a roughly 2:1 binary mass ratio and an age range of 2-6 Gyr. The observed temperatures, masses, and apparent radii match expected values from evolutionary models for a binary much better than a single-object model, but more detailed observations and modeling are needed to definitively confirm the binary hypothesis.

\end{abstract}

\section{Introduction}

GJ 229B is noted for being the first type-T brown dwarf to be discovered \citep{GJ229Disc}\footnote{Not to be confused with the Neptune-mass planet accompanying the host star, GJ 229Ab, sometimes written as GJ 229b \citep{GJ229Planet}.}, with a measured effective temperature well below the L/T transition, $\sim$1200~K. Yet its atmospheric structure and composition have been less thoroughly studied than now-more-standard benchmark objects such as the similar GJ 570D (e.g. \citealt{Line15}). This may be owing to its peculiar spectrum, which shows unusually weak H$_2$O and CH$_4$ features for its spectral type \citep{Burgasser06} and elevated levels of CO \citep{GJ229Keck}.

Dynamical measurements of the mass of GJ 229B have also yielded surprising results, with the most precise recent analysis finding a value of $71.4\pm0.6\,M_J$ \citep{Brandt}. This measurement was based on Gaia EDR3 data combined with several other astrometric and radial velocity data sets, resulting in an uncertainty significantly less than previous results. Yet it is incompatible with evolutionary models, as an object of that mass could not cool to the observed temperature of GJ 229B within a Hubble time. Meanwhile, an independent analysis of radial velocities alone found a minimum mass of only 1.62 $M_J$ \citep{Feng20}. Reconciling such a small minimum mass with the astrometric results requires fine-tuning the orbit solution with an orientation within $\sim$2$^\circ$ of face-on.

A newer analysis of the combined astrometric and RV data set was undertaken with the addition of the Gaia-Hipparcos positional difference by \cite{Feng22}. This analysis found a dynamical mass for GJ 229B of $60.42^{+2.34}_{-2.38}\,M_J$. This mass is somewhat more plausible in terms of fitting evolutionary models. However, it still strongly precludes fitting the observed dynamical mass, effective temperature, and luminosity (and by proxy radius) simultaneously.

Recently, we undertook a spectroscopic retrieval study of GJ 229B's atmosphere with our APOLLO atmospheric retrieval framework \citep{Howe22}, supplemented by an analysis incorporating the Sonora-Bobcat (S-B) grid of brown dwarf evolutionary models by \citet{Sonora}. 
This study showed that both spectroscopic retrieval measurements and evolutionary model fits return dramatically lower masses than the dynamical fits, with a retrieved mass estimate of $41.6\pm3.3\, M_J$. We also obtained an unexpectedly large radius estimate $1.105\pm0.025\,R_J$ (which is also implied by the luminosity and effective temperature of the object), which is inconsistent with evolutionary models of either mass, both of which predict a more compact object.

With such a large disparity in mass measurements, the most natural explanation would appear to be that GJ 229B is in fact a binary brown dwarf of unequal mass, with one component contributing most of the total flux to the spectrum, while the other contributes just enough flux to boost the apparent radius. This sort of discrepancy has been observed before for other objects, with binarity suggested as a solution to anomalously large retrieved radii \citep{Line17} and overall flux \citep{Burgasser08}, so a similar solution seems likely for GJ 229B. In this letter, we explore the possible parameter space for a binary GJ 229B using two different grid-fitting methods based on the S-B evolutionary models \citep{Sonora} and the APOLLO retrieval code, extending our analysis from \cite{Howe22}. First, we fit the observed spectrum of GJ 229B \citep{Geballe96,Noll97,Schultz98,GJ229Keck} to binary spectra derived from the S-B model grid directly. These atmosphere models are self-consistent, but do not account for the peculiar chemistry of GJ 229B or the super-Solar C/O ratio found by \cite{Howe22}. Second, we generate a grid of APOLLO forward models for a binary object based on values for radius, gravity, and temperature structure from the S-B evolutionary tracks, but using our retrieved molecular abundances from \cite{Howe22}. For completeness, we consider fits using both the $M_{\rm tot}=71\,M_J$ and $M_{\rm tot}=60\,M_J$ dynamical mass measurements.

For each of these grids, we compute goodness-of-fit statistics for comparison to the observed spectrum of GJ 229B, and we find that both grid fits return qualitatively similar results. In the $M_{\rm tot}=71\,M_J$ case, we find a primary mass of 51 $M_J$ ($q=0.39$) using both methods. The $M_{\rm tot}=60\,M_J$ case shows modest differences, with $M_1=46\,M_J$ ($q=0.30$) for S-B and $M_1=49\,M_J$ ($q=0.22$) for APOLLO. Given this consistency between our two methods and the physical arguments against a single-object solution, we consider this to be significant evidence for a binary solution for this object.

The results from the two model grid fits are shown in Section \ref{sec:evo}, with an explanation of the limitations of our model in Section \ref{sec:limits}. We discuss the implications of our results in Section \ref{sec:discuss} and summarize our conclusions in Section \ref{sec:conclusion}. A full description of the dataset we use in this paper is included in \cite{Howe22}.

\section{Grid Fits Based on Evolutionary Models}
\label{sec:evo}

We constructed two grids of spectra for binary brown dwarfs based on the solar-metallicity evolutionary tracks of the Sonora-Bobcat model set \citep{Sonora}. To create these spectra, we linearly interpolated the S-B tracks to a grid of 1 $M_J$ in mass and 10 K in primary effective temperature. For each grid point in the resulting ``S-B grid,'' we computed a primary spectrum and two secondary spectra by interpolating the nearest S-B spectra provided in the parameter space for two brown dwarfs of equal age and masses summing to 60 $M_J$ and 71 $M_J$. Likewise, for each grid point in the ``APOLLO grid,'' we computed forward models using APOLLO for the same radii, surface gravities, and temperature profiles from the S-B models combined with our retrieved molecular abundances from \cite{Howe22}. Thus, we used the same evolutionary tracks for both grids, but two different, incomplete, but complementary chemistry models.

For each grid and each case (60 and 71 $M_J$), we summed the primary and secondary spectra together to create a grid of model binary spectra. We then computed the goodness of fit to the observed spectrum of GJ 229B using a reduced chi-squared statistic. We chose the reduced chi-squared statistic for this analysis for its relative simplicity. However, most other measures of goodness of fit such as the Bayesian likelihood, Bayesian information criterion, and APOLLO's own likelihood function produce virtually identical results because all of these measures are dominated by the error in the large number of spectral data points (which comprise nearly all of the $>3200$ degrees of freedom) rather than other statistical factors.

\begin{figure}[h!]
    \centering
    \includegraphics[width=0.99\textwidth]{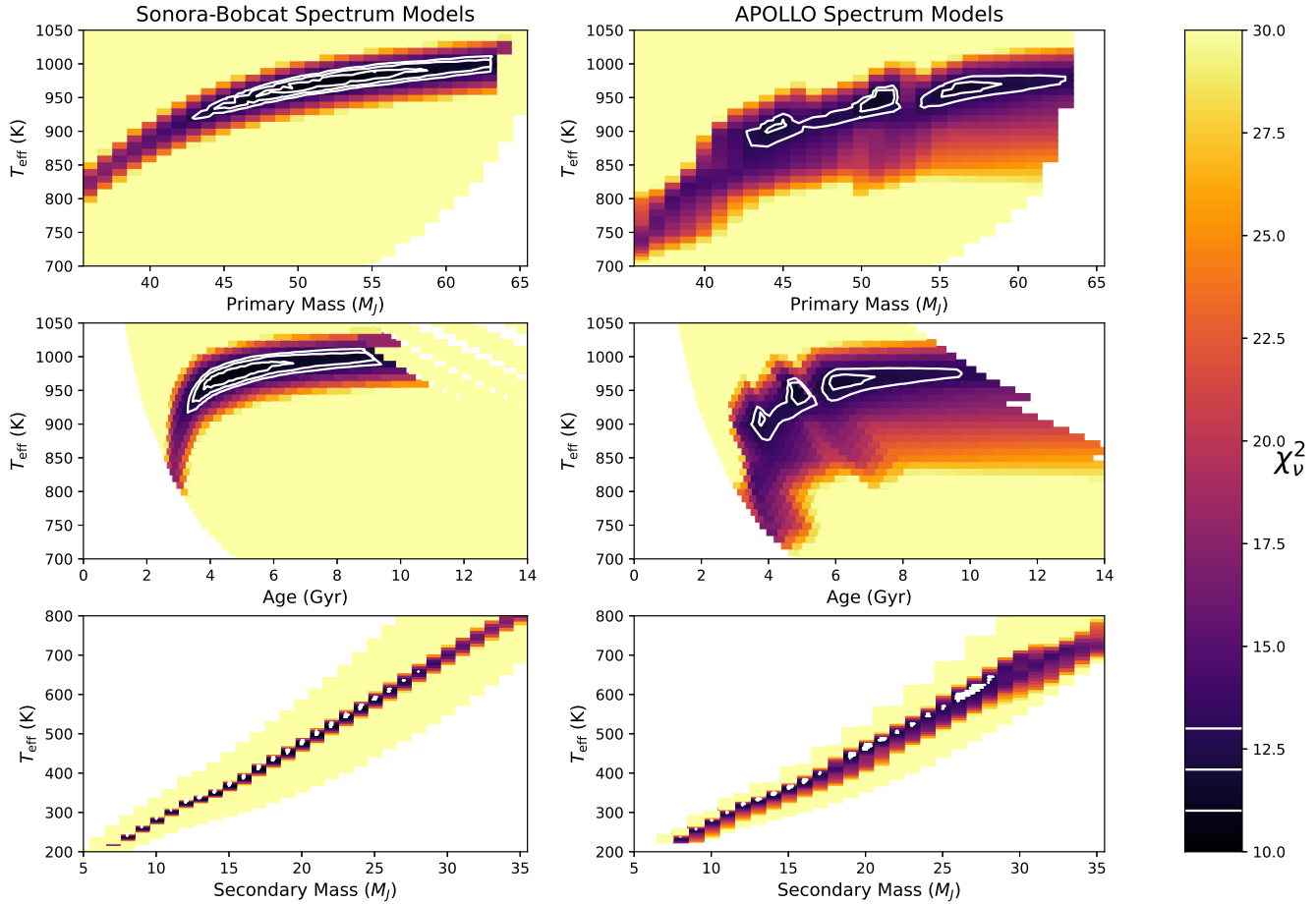}
    \caption{Goodness of fit to observed spectra of GJ 229B of our binary forward model grids, with a total mass of 71 $M_J$, calculated from the Sonora-Bobcat (S-B) models \citep{Sonora} (left) and from out APOLLO forward models (right). For the primary, the cutoff on the left edge is based on a mass greater than 50\% of the total mass of the system, and the cutoff on the right is based on an age less than a Hubble time. For the secondary, the low temperature cutoff is based on the limitations of the molecular cross section tables used by APOLLO. The outermost contours are representative of our adopted $1\sigma$ uncertainties, although they are slightly different for each local minimum.}
    \label{fig:heat}
\end{figure}

\begin{figure}[h!]
    \centering
    \includegraphics[width=0.99\textwidth]{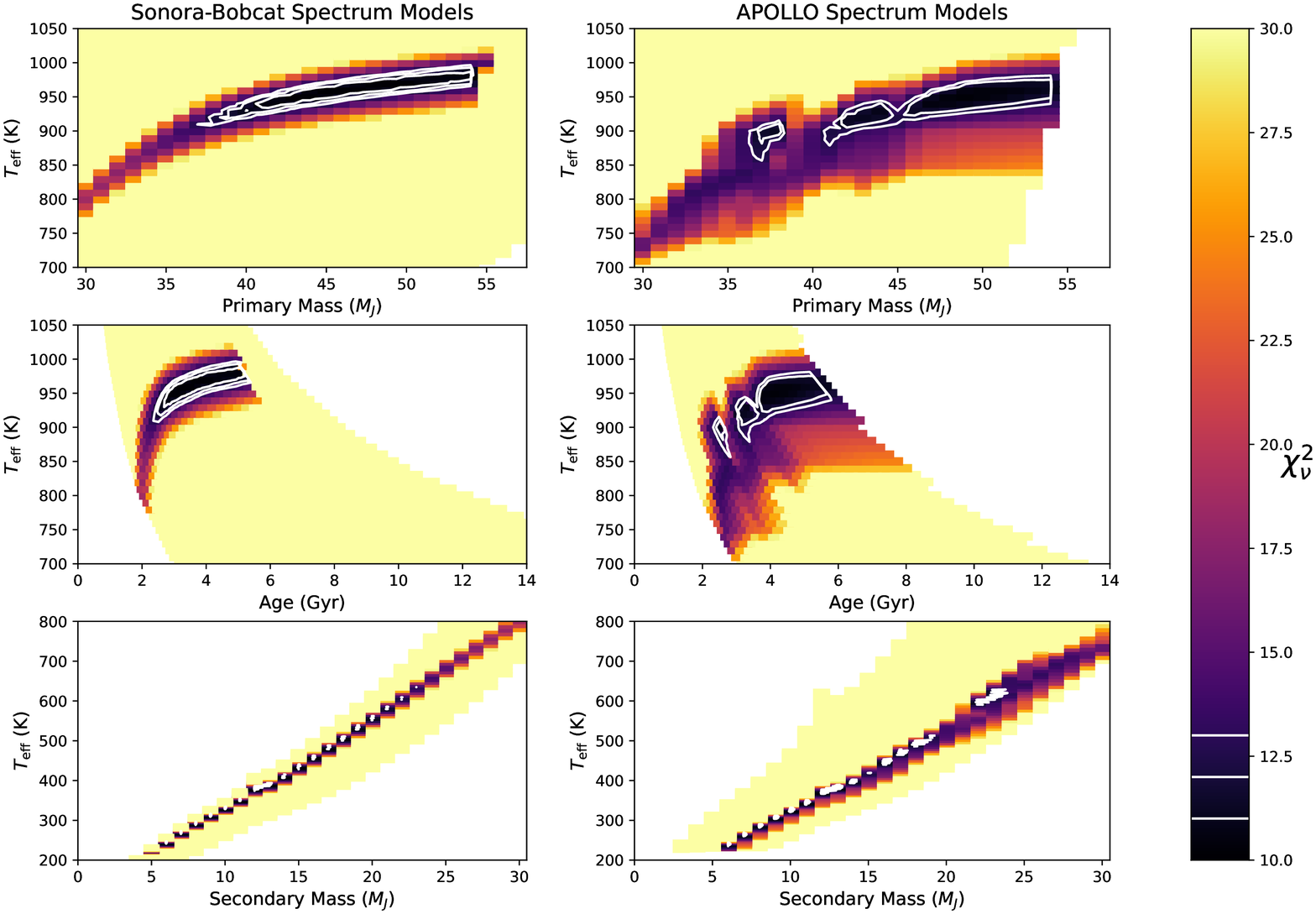}
    \caption{Same as Figure \ref{fig:heat}, except with a total mass of 60 $M_J$.}
    \label{fig:heat60}
\end{figure}

Heat maps of the reduced chi-squared results for both grids are shown in Figure \ref{fig:heat} for the $M_{\rm tot}=71\,M_J$ case and Figure \ref{fig:heat60} for the $M_{\rm tot}=60\,M_J$ case, plotting primary effective temperature versus mass and age, as well as secondary effective temperature versus mass. All of the fits were found to have $\chi_\nu^2\geq 10$, but this is not an unexpected result given that we are not performing full retrievals (such as MCMC retrievals) on the spectrum. To assess the uncertainties in these fits, we adopt $1\sigma$ errors of $\chi_\nu^2 \le 1.2\chi_{\nu,{\rm min}}^2$ based on an analysis of chi-squared statistics by \cite{StatsRef}. (See Figure 36.2 and associated chapter.) We mark reduced chi-squared contours on the plots of $\chi_\nu^2=11,\,12,\,{\rm and}\,13$. Based on our definition, the outer contour is roughly equivalent to our $1\sigma$ uncertainties, although they are slightly different for each local minimum.

All four grid fits result in a relatively broad minimum in $\chi_\nu^2$ spanning a width of $\sim$15 $M_J$. For example, for the S-B model grid, our $1\sigma$ uncertainties in primary mass span 44-63 $M_J$ in the 71 $M_J$ case and 39-54 $M_J$ in the 60 $M_J$ case. In all cases, the primary effective temperature falls in a significantly narrower range of $\sim$900-1000 K, which is consistent between the two total mass cases. Within each case, the S-B and APOLLO fits are qualitatively similar. However, the distribution for APOLLO forward models is broader in effective temperature and occurs at slightly lower temperatures overall. This is likely due to the chemistry model used for the APOLLO model grid, which assumes a constant atmospheric composition with respect to temperature, muting the effect of changing $T_{\rm eff}$ on the spectrum shape. Meanwhile, the realistic changes in chemistry in the S-B grid are likely to produce worse spectral fits at non-optimal temperatures, as would be expected.

The parameters of the best fits from our model grids are listed in Tables \ref{tab:models} and \ref{tab:models60} for the 71 $M_J$ and 60 $M_J$ cases, respectively. We note that while the S-B grid has a single global minimum in reduced chi-squared, in both of our cases, the APOLLO grid produces three local minima along a line of increasing mass and temperature. For completeness, we list all three of the APOLLO minima for each case in Tables \ref{tab:models} and \ref{tab:models60}. These are compared with our best single-object free retrieval from \cite{Howe22}, and reduced chi-squared values are listed for all of the fits. Including the three APOLLO fits is also useful in that they are representative of our $1\sigma$ uncertainty distributions as a whole.

In Table \ref{tab:models60}, we also include single-object fits to both of our grids, subject to the same dynamical mass constraint of 60$M_J$. These fits have higher reduced chi-squared values than the binary fits, so the binary solution is still favored in this case. For the 71 $M_J$ case, the global minimum in $\chi_\nu^2$ falls outside the distribution of allowed evolutionary models (and all models within the grid have very large $\chi_\nu^2>10000$), so we consider it to have no single-object solutions.


The most notable differences between the 60 $M_J$ and 71 $M_J$ cases are in the mass ratio of the binary and the inferred age of the system. The 9 $M_J$ difference in mass between the two cases is partitioned roughly equally between the primary and secondary mass, with the overall effect being a significantly more extreme mass ratio in the 60 $M_J$ case. The S-B best fit (which in both cases is also representative of the APOLLO best fits) in the 71 $M_J$ cass has a mass partition of 51 and 20 $M_J$ for a mass ratio of $q=0.392$. In the 60 $M_J$ case, the mass partition is 46 and 14 $M_J$, with $q=0.304$. Given a smaller total mass, a smaller primary mass with nearly the same $T_{\rm eff}$ has a larger radius and thus produces a larger fraction of the total flux of the binary, so a cooler and lower-mass secondary is needed to make up the difference.

Likewise, a lower mass primary would need a shorter time to cool to the same $T_{\rm eff}$, resulting in a younger inferred age of the system. For the S-B best fit, the difference is about 1 Gyr, reducing from 4.5 Gyr in the 71 $M_J$ case to 3.5 Gyr in the 60 $M_J$ case. All of the fits listed in Tables \ref{tab:models} and \ref{tab:models60} fall in the range of 2.6-6.1 Gyr, somewhat older than previous estimates (e.g. \cite{GJ229Age}).

\begin{table}[htb]
    \hspace{-0.9in}
    \begin{tabular}{l | r | r | r | r | r | r | r}
    \hline
         & \multicolumn{4}{c|}{Binary} & \multicolumn{3}{c}{Single} \\
    \hline
         & S-B & APOLLO 1 & APOLLO 2 & APOLLO 3 & S-B & APOLLO & Free Retrieval \\
    \hline
    \hline
    Primary Mass ($M_J$)        & $51_{-7}^{+12}$           & $45_{-3}^{+8}$            & $51_{-8}^{+1}$            & $56_{-2}^{+6}$            & No Solution & No Solution &  $41.6\pm3.3$ \\
    Primary $T_{\rm eff}$ (K)   & $970_{-50}^{+40}$         & $910\pm60$                & $950_{-80}^{+20}$         & $960\pm30$                & \ & \ &  $869_{-7}^{+5}$ \\
    Primary Radius ($R_J$)      & $0.811_{-0.045}^{+0.040}$ & $0.833_{-0.033}^{+0.009}$ & $0.809_{-0.007}^{+0.033}$ & $0.788_{-0.026}^{+0.008}$ & \ & \ &  $1.105\pm0.025$ \\
    Primary ${\rm log}\,g$      & $5.30_{-0.12}^{+0.15}$    & $5.23_{-0.05}^{+0.09}$    & $5.31_{-0.11}^{+0.01}$    & $5.37_{-0.02}^{+0.08}$    & \ & \ &  $4.93_{-0.03}^{+0.02}$ \\
    \hline
    Secondary Mass ($M_J$)      & $20_{-12}^{+7}$           & $26_{-8}^{+3}$            & $20_{-1}^{+8}$            & $15_{-6}^{+2}$            & \ & \ & \\
    Secondary $T_{\rm eff}$ (K) & $478_{-216}^{+210}$       & $596_{-164}^{+77}$        & $469_{-32}^{+176}$        & $364_{-130}^{+38}$        & \ & \ &  \\
    Secondary Radius ($R_J$)    & $0.940_{-0.039}^{+0.069}$ & $0.909_{-0.013}^{+0.036}$ & $0.938_{-0.040}^{+0.006}$ & $0.972_{-0.019}^{+0.036}$ & \ & \ &  \\
    Secondary ${\rm log}\,g$    & $4.77_{-0.46}^{+0.20}$    & $4.91_{-0.17}^{+0.08}$    & $4.77_{-0.03}^{+0.18}$    & $4.61_{-0.30}^{+0.08}$    & \ & \ &  \\
    \hline
    Mass Ratio ($M_2/M_1$)      & $0.392_{-0.265}^{+0.221}$    & $0.578_{-0.238}^{+0.212}$ & $0.392_{-0.027}^{+0.298}$ & $0.268_{-0.122}^{+0.047}$ & \ & \ &  \\
    Flux Ratio ($F_2/F_1$)      & $0.079_{-0.070}^{+0.224}$ & $0.219_{-0.135}^{+0.073}$ & $0.080_{-0.001}^{+0.166}$ & $0.031_{-0.025}^{+0.012}$ & \ & \ &  \\
    Age (Gyr)                   & $4.5_{-1.4}^{+4.7}$       & $3.8_{-0.5}^{+1.8}$       & $4.8_{-1.4}^{+0.7}$       & $6.1_{-0.6}^{+4.0}$       & \ & \ &  $>$1.0 \\
    $\chi_\nu^2$                & 10.37 & 11.81 & 11.14 & 11.54 & \ & \ &  8.15 \\
    \hline
    \end{tabular}
    \caption{The best fit binary models from the Sonora-Bobcat (S-B) model set and our APOLLO forward model grid with a total mass of 71 $M_J$. Bolometric flux ratios are calculated from the Stephan-Boltzmann law. The retrieved parameters are compared with our best fit free retrieval for a single object \citep{Howe22}. No solutions for a single object with a global $\chi_\nu^2$ minimum lie within either of the model grids, which we interpret as having no solution with physically plausible parameters. Uncertainties are defined by $\chi_\nu^2 \le 1.2\chi_{\nu,{\rm min}}^2$ based on an analysis by \cite{StatsRef}. \\ $^a$The APOLLO 1 and APOLLO 2 models are two peaks of a shared $1\sigma$ distribution.}
    \label{tab:models}
\end{table}

\begin{table}[htb]
    \hspace{-0.9in}
    \begin{tabular}{l | r | r | r | r | r | r | r}
    \hline
         & \multicolumn{4}{c|}{Binary} & \multicolumn{3}{c}{Single} \\
    \hline
         & S-B & APOLLO 1 & APOLLO 2 & APOLLO 3 & S-B & APOLLO & Free Retrieval \\
    \hline
    \hline
    Primary Mass ($M_J$)        & $46_{-7}^{+8}$            & $38_{-2}^{+0.5}$          & $44_{-3}^{+1}$            & $49_{-3}^{+5}$            &    60                     &    60                     & $41.6\pm3.3$ \\
    Primary $T_{\rm eff}$ (K)   & $960_{-40}^{+30}$         & $900_{-50}^{+20}$         & $930_{-50}^{+10}$         & $950\pm30$                & $1000_{-20}^{+10}$        & $970\pm20$                & $869_{-7}^{+5}$ \\
    Primary Radius ($R_J$)      & $0.835_{-0.033}^{+0.061}$ & $0.872_{-0.002}^{+0.011}$ & $0.841_{-0.003}^{+0.013}$ & $0.818_{-0.023}^{+0.014}$ & $0.777_{-0.002}^{+0.001}$ & $0.773_{-0.002}^{+0.003}$ & $1.105\pm0.025$ \\
    Primary ${\rm log}\,g$      & $5.23_{-0.18}^{+0.13}$    & $5.11_{-0.08}^{+0.01}$    & $5.21_{-0.05}^{+0.01}$    & $5.28_{-0.04}^{+0.06}$    & $5.412\pm0.002$           & $5.416\pm0.003$           & $4.93_{-0.03}^{+0.02}$ \\
    \hline
    Secondary Mass ($M_J$)      & $14_{-8}^{+7}$            & $22_{-0.5}^{+2}$          & $16_{-1}^{+3}$            & $11_{-5}^{+3}$            & \ & \ & \\
    Secondary $T_{\rm eff}$ (K) & $411_{-169}^{+174}$       & $597_{-7}^{+38}$          & $449_{-30}^{+64}$         & $346_{-108}^{+56}$        & \ & \ & \\
    Secondary Radius ($R_J$)    & $0.990_{-0.051}^{+0.059}$ & $0.944_{-0.018}^{+0.002}$ & $0.975_{-0.021}^{+0.002}$ & $1.011_{-0.024}^{+0.028}$ & \ & \ & \\
    Secondary ${\rm log}\,g$    & $4.57_{-0.50}^{+0.30}$    & $4.81_{-0.01}^{+0.11}$    & $4.64_{-0.07}^{+0.09}$    & $4.45_{-0.29}^{+0.12}$    & \ & \ & \\
    \hline
    Mass Ratio ($M_2/M_1$)      & $0.304_{-0.193}^{+0.234}$ & $0.579_{-0.021}^{+0.088}$ & $0.364_{-0.030}^{+0.100}$ & $0.224_{-0.113}^{+0.080}$ & \ & \ & \\
    Flux Ratio ($F_2/F_1$)      & $0.047_{-0.031}^{+0.120}$ & $0.227_{-0.001}^{+0.036}$ & $0.073_{-0.006}^{+0.116}$ & $0.027_{-0.020}^{+0.017}$ & \ & \ & \\
    Age (Gyr)                   & $3.5_{-1.0}^{+1.8}$       & $2.6\pm0.3$               & $3.4_{-0.4}^{+0.2}$       & $4.3_{-0.6}^{+1.4}$       & $7.0_{-0.2}^{+0.4}$       & $7.7_{-0.4}^{+0.5}$       & $>$1.0 \\
    $\chi_\nu^2$                &  9.97 & 11.22 & 10.92 & 10.25 & 11.25 & 24.56 & 8.15 \\
    \hline
    \end{tabular}
    \caption{Same as Table \ref{tab:models}, except with a total mass of 60 $M_J$. Single-object solutions for both grid fits are listed alongside the free retrieval from \cite{Howe22}.}
    \label{tab:models60}
\end{table}

\begin{figure}[h!]
    \centering
    \includegraphics[width=0.99\textwidth]{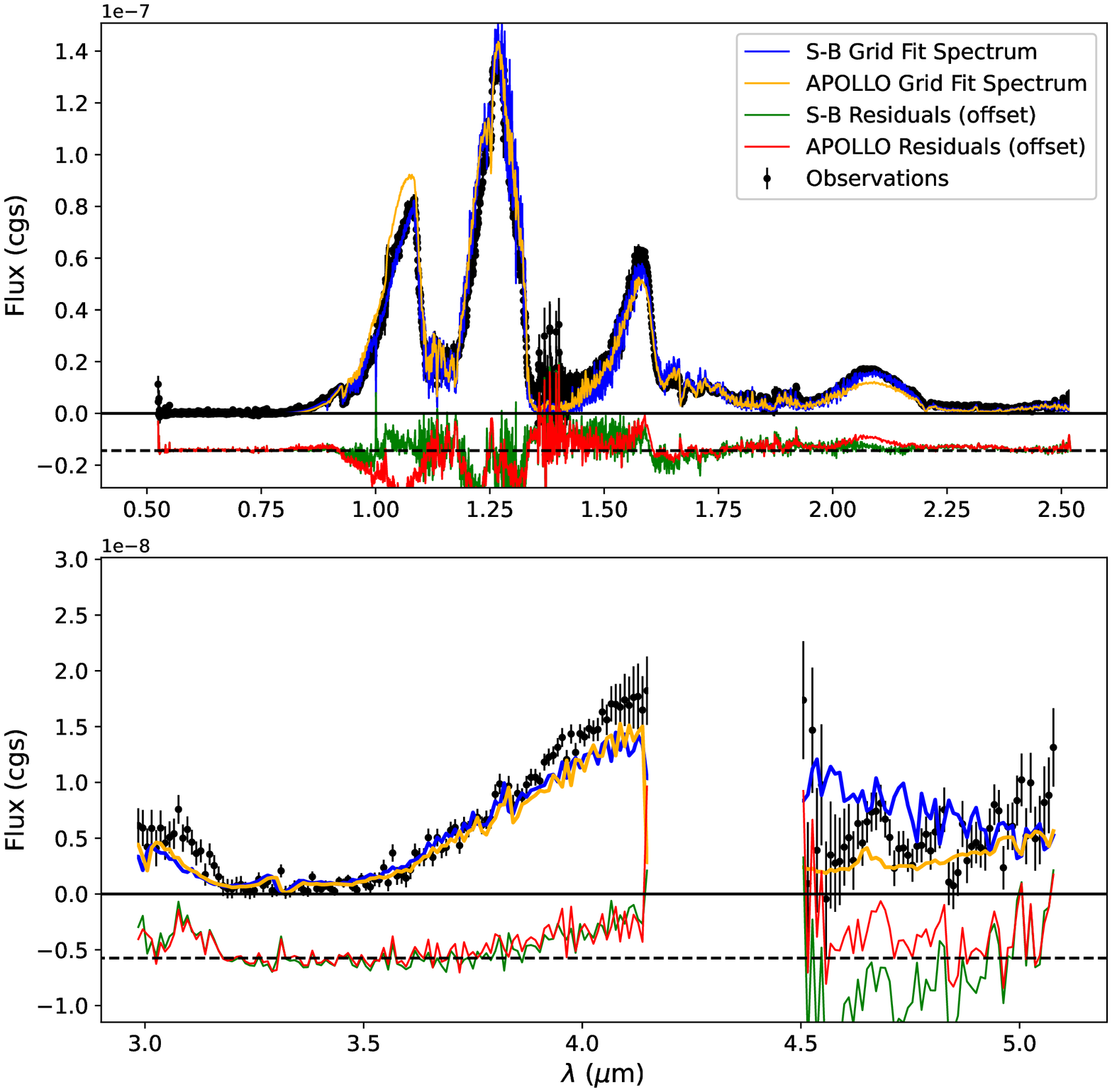}
    \caption{Comparison of the observed spectrum of GJ 229B (black) with best fit binary spectra from the S-B model grid (blue) and our APOLLO forward model grid (gold), in the 71 $M_J$ case. Residuals for each fit are plotted on the same scale below the x-axis in green and red, respectively.}
    \label{fig:gridspec}
\end{figure}

As a representative example of the spectra resulting from our grid fits, Figure \ref{fig:gridspec} compares the best fit spectra from each grid (blue and gold) and their residuals (green and red) with the observed spectrum of GJ 229B in the 71 $M_J$ case. Both grid fits match the observed spectrum at many wavelengths; however, they both show errors in different wavelength ranges. The S-B fit shows significant residuals in the J- and M-bands, while the APOLLO fit shows significant residuals mainly in the Y- and K-bands. Note that features in the residual spectra are exaggerated in regions of low observational uncertainties, so their magnitude is more indicative of the observational precision than actual absorption features.

\section{Limitations of the Model}
\label{sec:limits}

The Sonora-Bobcat (S-B) model set provides a complete grid of spectra only for brown dwarf models of solar metallicity and C/O ratio. Therefore, a detailed grid fit to an observed spectrum must make compromises with regard to the chemistry. Our S-B grid fit to GJ 229B assumes an essentially Solar composition, which is inconsistent with both the peculiar spectrum of this object and the retrieved molecular abundances from \cite{Howe22}. Our measured super-Solar C/O ratio of 1.13 is consistent with other retrievals of T dwarfs (e.g. \cite{Line15}), but is well outside the {\it a priori} parameters of generic brown dwarf model sets such as Sonora-Bobcat.

Additionally, the temperature-pressure profiles in both forward model grids are interpolated from the S-B models. They are not informed by our retrieved T-P profile from \cite{Howe22}, and in the context of the grid, it is {\it de facto} a two-parameter model controlled only by $T_{\rm eff}$ and log$\,g$. This is analogous to the parametric T-P profile used in \cite{Howe22}, which we found to return relatively poor goodness-of-fit statistics, even in a full ensemble retrieval.

Meanwhile, our APOLLO grid assumes a constant molecular composition for brown dwarfs of different temperatures, as a full thermochemical model of the atmosphere is beyond the scope of APOLLO, which is designed around a free retrieval philosophy for a single object. Therefore, neither of the model grids used in this analysis fully simulate the atmospheric chemistry of GJ 229B. A fully self-consistent chemical model with super-Solar C/O would be needed to model the spectrum of a binary brown dwarf to a greater accuracy than we have done in this paper.

The limitations of the chemical model for the APOLLO grid are especially significant for the secondary spectrum, which is much cooler than the effective temperature retrieved for a single-object fit. Thus, the atmospheric chemistry is expected to be quite different from our assumed abundances. In performing the fits with the APOLLO grid, we assume that the relatively smaller flux from the secondary means that it will produce a proportionately smaller error in the overall spectrum.

An additional limitation of the APOLLO-based grid was the inability of APOLLO to model atmospheres of very cold (old and low-mass) secondaries due to the low-temperature limits of our molecular optical cross-section tables. This prevented us from fully exploring the parameter space at ages $\gtrsim10$ Gyr in the 71 $M_J$ case and $\gtrsim6$ Gyr in the 60 $M_J$. However, the contour plots in Figures \ref{fig:heat} and \ref{fig:heat60} indicate that the best-fit models are clustered at younger ages in the region of the parameter space we did explore.

The reduced chi-squared values for our grid fits in this work are consistently significantly larger than that of our free retrieval of the single-object solution for GJ 229B in \cite{Howe22}. We attribute this to our less accurate treatment of the chemistry in both model grids, as described above, and to the simplified T-P profiles. The goodness-of-fit when performing a free retrieval using a parametric T-P profile was comparable to that of our grid fits in this work, which may be a result of both models being underdetermined to fully fit the observed spectrum.

We do not attempt a full retrieval of the prospective binary spectrum in this work, as would be required to make an accurate comparison of the goodness-of-fit between the single and binary models. The number of variables involved in retrieving properties of two independent objects from a blended spectrum is prohibitive and beyond the scope of the current version of APOLLO, and a free retrieval on a blended spectrum would likely have too many degeneracies to produce unambiguous results. Again, a full chemical model for a binary object would be needed to constrain the parameter space.

Nonetheless, we see that both of our grid fits produce similar results for best fits in $M$-$T_{\rm eff}$-space, and they also fit complementary wavelength regimes of the observed spectrum. The consistency of this result across two methods emphasizing different chemistry increases our confidence in the viability of our method to infer the properties of a binary brown dwarf in this context.

\section{Discussion}
\label{sec:discuss}

\subsection{Physical Argument for a Binary Solution}
\label{sec:single}

Our investigation of binarity for GJ 229B was motivated by multiple lines of evidence that render a single-object solution improbable, if not outright unphysical, given our current understanding of brown dwarf evolution. First, the combination of observed luminosity and effective temperature, even prior to dynamical mass measurements, implies an unusually large radius for a single object. Our single-object retrieval in \cite{Howe22} found a radius of 1.105$\pm$0.025 $R_J$. When combined with our retrieved $T_{\rm eff}=869_{-7}^{+5}$ K, based on the S-B models, this suggests a very young and low mass object with an age of $<0.5$ Gyr and a mass near the deuterium-burning limit.


While this solution could fit the luminosity and effective temperature of GJ 229B, it is wildly inconsistent with the very large dynamical mass measurements made in recent years. A massive brown dwarf of 71 $M_J$ would not be able to cool to the observed temperature of GJ 229B in a Hubble time, and the S-B evolutionary tracks include no valid solutions for GJ 229B with this mass. 
A 60 $M_J$ brown dwarf could cool sufficiently in $\sim$7-10 Gyr. However, these solutions still fail with high confidence to replicate our retrieved radius, setting an upper bound of $R\le0.77\,R_J$. In contrast, a binary object with an unequal flux ratio would replicate the apparent oversized-radius much more easily.

\subsection{Inferred Mass Ratio of GJ 229B in the Context of Other Brown Dwarf Binaries}

Our inferred mass ratio for a potential binary GJ 229B is notable in that it is fairly unequal, with a best-fit solution of $q=0.39$ in the 71 $M_J$ case for both grid fits and an even more extreme $q=0.30$ in the 60 $M_J$ case for the S-B fit (albeit with large uncertainties). In contrast, the most notable well-constrained dynamical mass ratios measured for binary brown dwarfs are all near $q\sim0.8$ \citep{EIndiMass,Luhman16,GJ569}. However, inferred mass ratios for other binaries based on evolutionary models are more varied, with several estimated to be near $q\sim0.5$ \citep{WISE2150,Gonzales20,Reiners10}.

Population-level studies of brown dwarf binaries are subject to greater systematic uncertainties because dynamical masses are generally not available, forcing them to rely on evolutionary models and mass-luminosity relations. Early population studies along these lines yielded conflicting results. \cite{Burgasser03} did a survey of T dwarf binaries and found significant numbers of systems down to $q\sim0.4$, based on a mass-luminosity relation. However, in a later study, \cite{BurgasserBin} estimated 77\% of very-low-mass (VLM, $<0.1\,M_\odot$) binaries have $q\ge0.8$; but this study reported low completeness for unequal ratios of $q\le0.6$, and their best-fit model suggested that mass ratios as low as $q\sim0.4$ may be plausible for VLM binaries. In a later study, \cite{Dupuy17} measures mass ratios for 19 binary brown dwarfs and found that most of them had $q\gtrsim0.8$, and all of them had $q\gtrsim0.6$.

These population studies do not show strong evidence for highly unequal binary brown dwarf mass ratios. However, some of them do extrapolate the binary brown dwarf population to be consistent with a mass ratio for GJ 229B of $q=0.39$. More promising is that the uncertainties in our analysis allow a much more equal mass ratio, reaching as high as $q=0.79$ for the APOLLO 1 fit in the 71 $M_J$ case. This indicates that there are a significant number of solutions that are consistent with population studies, so this is not a large obstacle to a binary interpretation for GJ 229B.

\subsection{Detectability of a Binary}

The definitive proof of binarity for this object would of course be direct observation, either by resolving the separate components, or by observing the reflex motion of the primary. At least one and possibly both of these methods should be able to significantly constrain the parameters of the binary. At a distance of 5.76 pc, the fact that the two components are not resolved in {\it HST} images \citep{GJ229Disc} suggests that their projected separation is $<0.5$ AU. However, we note that adaptive optics measurements with ground-based telescopes stand a fair chance of being able to resolve the components or measure the motion of the centroid. Typical adaptive optics systems will offer a resolution as small as $\sim$0.2 AU at that distance, and the GRAVITY instrument on VLT may be able to resolve an order of magnitude closer \citep{VLT}. As for radial velocity characterization, the apparent face-on orientation of the system as a whole as derived from astrometry suggests that any signal may be weak. However, if the orbit of the binary is significantly tilted relative to their mutual orbit around GJ 229A, their orbital motion at such small separations (at least several km s$^{-1}$ for the primary) should be well within the capability of ground-based spectroscopy.

For both of the most recent dynamical mass measurements of GJ 229B, none of the S-B evolutionary models can replicate the dynamical mass, effective temperature, and luminosity simultaneously for a single object, whereas a binary solution does so quite easily. Therefore, we feel confident in proposing it as a viable solution to the puzzles surrounding this object. We further note that our binary fits consistently predict an older age range for the system of 2-6 Gyr. In the absence of direct dynamical or spectroscopic evidence for binarity, an asteroseimic age measurement of GJ 229A could help to further constraint the parameter space for a binary brown dwarf companion.

\section{Conclusions}
\label{sec:conclusion}

Both spectroscopic retrievals and SED considerations combined with evolutionary models consistently suggest a smaller mass for the late T-dwarf GJ 229B than the dynamical mass of $71.4\pm0.6\,M_J$ \citep{Brandt}, or $60.42^{+2.34}_{-2.38}\,M_J$ \citep{Feng22}, suggesting that it may be a binary object. We have employed grid-based methods using the Sonora-Bobcat evolutionary models \citep{Sonora} and the APOLLO retrieval code \citep{Howe22} to estimate the mass ratio of a prospective binary GJ 229Ba and Bb. Our best fit solutions for the S-B grid fits (which have the smallest reduced chi-squared values) give a mass ratio of $0.39_{-0.27}^{+0.22}$ for the 71 $M_J$ case and $0.30_{-0.19}^{+0.23}$ for the 60 $M_J$ case. While these ratios are fairly unequal, they do overlap with the ranges inferred for resolved brown dwarf binaries from some population studies. These fits are also consistent with an intermediate age range for the system of 2-6 Gyr.

There is a significant probability that precise astrometric and/or radial velocity measurements will be able to confirm the binarity of GJ 229B. In the absence of such measurements, a blended spectrum retrieval incorporating an equilibrium chemistry model may be able to significantly refine the mass ratio estimated in this work.




\acknowledgements

ARH was supported by an appointment to the NASA Postdoctoral Program at NASA Goddard Space Flight Center, administered by Oak Ridge Associated Universities under contract with NASA. ARH also acknowledges support by NASA under award number 80GSFC21M0002 through the CRESST II cooperative agreement. AMM acknowledges support from GSFC Sellers Exoplanet Environments Collaboration (SEEC), which is funded in part by the NASA Planetary Science Division’s Internal Scientist Funding Model. This work was partially supported by the GSFC Exoplanets Spectroscopy Technologies (ExoSpec), which is part of the NASA Astrophysics Science Division's Internal Scientist Funding Model. We thank Beth Biller and Robert Haring-Kaye for helpful conversations.

\bibliography{refs}{}
\bibliographystyle{aasjournal}

\end{document}